# Steering of high energy electron beam in laser plasma accelerators

**Chang-qing Zhu,**[1,3] **Jin-guang Wang,**[1,3] **Yi-fei Li,**[1,3] **Jie Feng,**[1,3] **Da-zhang Li,**[4] **Yu-hang He,**[1,3] **Jun-hao Tan,**[1,3] **Jing-long Ma,**[1] **Xin Lu,**[1,3] **Yu-tong Li,**[1, 2, 3, 5]  **Li-ming Chen,**[2, 3, *]

[1]*Beijing National Laboratory of Condensed Matter Physics, Institute of Physics, CAS, Beijing 100190, China*
[2]*IFSA Collaborative Innovation Center and Department of Physics and Astronomy, Shanghai Jiao Tong University, Shanghai 200240, China*
[3]*School of physical Sciences, University of Chinese Academy of Sciences, Beijing 100049, China*
[4]*Institute of High Energy Physics, CAS, Beijing 100049, People's Republic of China*
[5]*Songshan Lake Materials Laboratory, Dongguan, Guangdong 523808, China*
*\* lmchen@iphy.ac.cn*

By using Dazzler system and tilting compressor grating, we provide an effective way of using the laser group delay dispersion (GDD) to continuously steer the high energy electron beam which is accelerated by asymmetric laser-wakefield. The deviation angle of electrons is as the same as the angular chirped laser pulse from its initial optical axis, which is determined by the laser pulse-front-tilt (PFT). This unique method can be continuously used to control over the pointing direction of electron-pulses to the requisite trajectories, especially for the alignment sensitive devices such as electron-positron collider or undulator. Besides, the effect of PFT on the qualities of electron beam has been investigated.

## 1. Introduction

Laser wakefield acceleration (LWFA) has been developed many years [1] and has made great progress based on chirped-pulse-amplification technology [2]. The ultrashort laser pulse with ultrahigh peak power can drive nonlinear plasma waves [3] of which the longitudinal electric fields can be very strong to accelerate electrons to several GeV [4,5]. The peak current of electron beams has been generated to tens of kA [6,7]. The qualities of produced electrons, including pointing stability, reproducibility and tunability, are significant for practical applications. For instance, the pointing direction of electron beams is vital for the users of electron-positron colliders or undulators [8-10]. The all-optical way by rotating the compressor grating to control over the pointing direction of LWFA electron-pulses was first experimentally detected by Popp et al. [11]. However, it is impractical for users to realign the laser beam path once the compressor grating is tilted. The electrons cannot be steered continuously. Another experimental scheme proposed by Nakanii et al. [12] to control the electrons with a static tilted magnetic field is designed ingeniously. However, the electron energy is limited to only about 10 MeV because of the density threshold. At low plasma density, the refraction effect stops working. It is noteworthy that the low plasma density is necessary for generating high-energy electrons in laser wakefield [13]. So, continuously guiding energetic electron beam with low energy spread to undulator without changing optical alignment is still a big problem. On the other hand, the angular misalignments in the double pass compressor used to temporally compress the pulses [2] are unavoidable, which can result in spatio-temporal distortions of the laser pulse and ascribe to deflect the accelerated electron beam. However, these induced distortions will deteriorate temporal resolution and reduce laser intensity etc [14], leading to the sacrifice of qualities of LWFA-electrons.

The grating misalignment will cause an unwanted angular chirp (AC) in the pulse near field [11]. If we consider a small deviation $\delta$ between the two gratings parallel to its grooves, the AC can be calculated as $\beta=d\theta/d\lambda=2\delta(tan\varepsilon)/(scos\alpha)$ [15]. Here $\alpha$, $\varepsilon$ and $s$ are the angle of incidence, diffraction angle of single grating and the groove spacing respectively. In general, the PFT [16] $p$ is consisted of two terms: $p=p_{AC}+p_{SC+GDD}$, where the first term $p_{AC} = -2\pi\beta/\omega$ is introduced by the angular chirp, while the second term $p_{SC+GDD}=\varphi^{(2)}\upsilon$ is induced by the spatial chirp $\zeta=dx_0/d\omega$ which is characterized by the frequency gradient $\upsilon$, and laser GDD $\varphi^{(2)}$ [14]. In LWFA, the laser pulse with PFT will excite asymmetric wakefield which will force the pulse to deviate from its initial axis and alter the way the electrons are injected and accelerated [11,17,18]. In the community, it is well known that the influence of GDD on the electron accelerating process has been studied and many groups use the algorithms to manage electron beam optimization [19,20]. However, the controlling over the pointing direction of high energy electron beams by GDD has not been reported yet.

In this letter, we study the effect of GDD on the pointing-directions and qualities of high energy electron beam accelerated by angularly chirped laser excited wakefield. The deviations are completely determined by the laser PFT. In the ionization injection regime, the asymmetric wake deflects the electrons from the initial optical axis. And also, even the introduced spatio-temporal distortions will erode the laser intensity which would negatively affect electron acceleration. While, with suitable GDD, the pulse duration can be corrected and the electrons can be optimized for different amounts of $\beta$.

## 2. Experimental setup

The Pulsar CPA laser system delivers 25 fs, 800-nm laser pulse with energy of ~400 mJ on pure-nitrogen gas target to produce the stable electrons. The p-polarized pulse focused by an F/12.5 gold coated off-axis parabola mirror had a spot size of $w_0$=9.5μm. The experimental setup is shown in Fig. 1, as characterized in Ref [21]. Without AC, the pointing stability (RMS) of continuous 30 shots of electron beams was ~0.5mrad, and the background electron density of every shot was ~$8.0\times10^{18}$cm$^{-3}$ while electron injection occurred. If the AC exists, the overall beam width will increase from $w_0$ due to spatial chirp $\zeta(z)= \zeta_0-\lambda\beta z/\omega$, which is seen in Fig. 1(b). Besides, the changed temporal chirp can reduce the spectral bandwidth locally and lengthen the pulse duration due to GDD $\varphi_0^{(2)}(z)= \varphi_0^{(2)}-2\pi\lambda\beta^2z/\omega^2$ [14]. Here, $z$ is the position of ultrashort-pulse in the propagation direction.

In the Pulsar, the Dazzler (or acousto-optic programmable dispersive filter) [22] is used to manipulate the spectral phase, $\varphi(\omega)$, of a laser pulse. One of the spectral phase term, $\partial^2\phi(\omega_0)/\partial\omega^2$, the GDD, is utilized to describe the linear chirp [20,23]. The spectral phase includes many high-order dispersions. Here, we investigate the effect of GDD with angularly chirped laser pulses on LWFA-electrons.

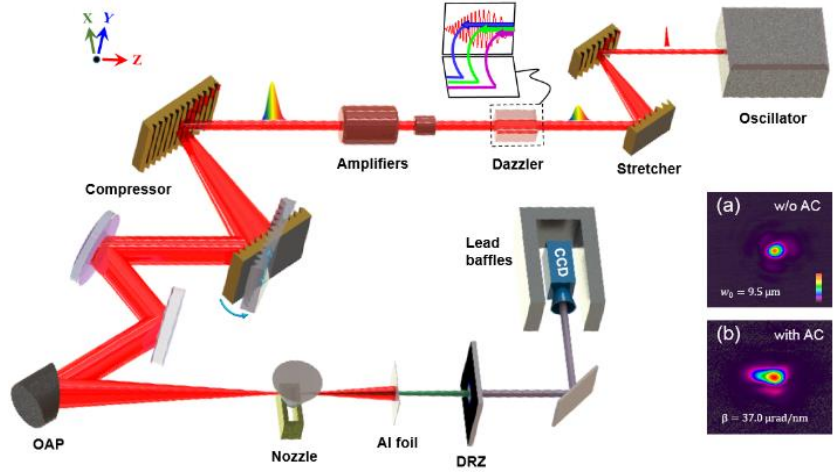

**FIG. 1.** Experimental setup. The insert (a) and (b) shows the laser focal spot without AC and with AC of β=37.0 μrad/nm respectively.

## 3. Results and discussions

When one of the two compressor gratings is rotated around the axis parallel to the groove, the angularly chirped or pulse-front tilted laser beam will be introduced. As a result, the light path might be altered. Thus, it needs to be realigned by turning the last mirror in the compressor.

The different spectral phase shift $\Delta\phi(\lambda,x)$ in transverse dimension $x$ is created by AC and leads to the position-dependent laser components of linear phase chirp $\Delta\phi(\lambda,x)/d\lambda$ with the perpendicular group delay $\Delta\phi(\omega,x)/d\omega$, where $\lambda_0$ is the laser central wavelength, $\omega_0=2\pi c/\lambda_0$ and $c$ the light speed in vacuum. Consequently, the laser intensity front is tilted with respect to the laser phase fronts by the angle $\psi=\lambda_0\beta$ [15], which is seen in Fig. 2(a). The Thomson scattered images from top-view clearly demonstrate that the angularly chirped laser pulse is deflected from its initial optical axis in plasma, shown in Fig. 2(b), (c) and (d). And the tilted angles $\psi$ are -4.74, 0.59, and 2.37 mrad, respectively. Obviously, the laser is deflected upwards (downwards) while $\beta$ is negative (positive). And the deviation angle $\gamma_1$ and $\gamma_2$ are -16.40 mrad and 14.65 mrad respectively, as shown in Fig. 2(b) and (d). Here the minus represents upwards. The electron beam relative to Fig. 2(b) and (d) leaves the initial laser propagation direction under angles of -15.58 mrad and 13.08 mrad respectively, which are determined by the directions of asymmetric wakefield, as shown in Fig. 3(a.I) and (a.III). While, for small β, the deflection of laser beam is inconspicuous, as also to the electron beams, seen in Fig. 2(c) and Fig. 3(a.II).

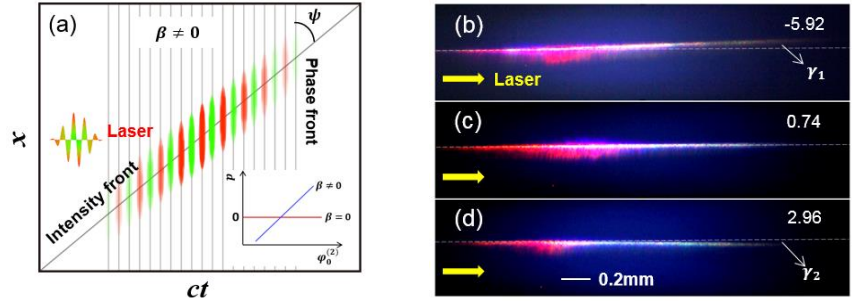

**FIG. 2.** (a) The laser intensity front tilted with respect to the laser phase fronts. The lower right insert shows the relationship between $p$ and $\beta$, $\varphi_0^{(2)}$. Thomson scattered images from top-view for different $\beta$: (b) -5.92, (c) 0.74 and (d) 2.96 [μrad/nm]. The dashed white line represents the initial laser axis. $\gamma_1$ and $\gamma_2$ stands for the deviation angles of laser from its initial direction.

When the ultrashort laser pulse propagate in the compressor with angular dispersion, the PFT $p$ can be characterized by [14]

$$p = -\frac{2\pi\beta}{\omega} + \varphi^{(2)}(z)\frac{\varsigma(z)}{\varsigma^2(z) + \frac{w_0^2 \tau_0^2}{4}}, \quad (1)$$

where $\tau_0$ is the pulse duration. From Eq. (1), it is easy to find the relationship between $p$ and $\varphi_0^{(2)}$ for different $\beta$, as depicted in the insert of Fig. 2(a). If the PFT is introduced, it will lead to the transverse plasma density gradient and variation of the refractive index perpendicular to the initial optical axis. Thus the intensity-front-tilted laser pulse will be pushed transversely and leads to the deflection of wakefield's from its propagation direction. The longitudinal density gradient has no significant change. The longitudinal electric field will accelerate the off-axis-injected electrons, while the transverse electric field will drive the electrons towards the bubble center axis [11]. Although the presence of PFTs in the laser pulse erodes the pulse intensity, it facilitates off-axis electron injection into the plasma wakefield. In a word, the deviation angle of accelerated electron is completely determined by the laser PFT.

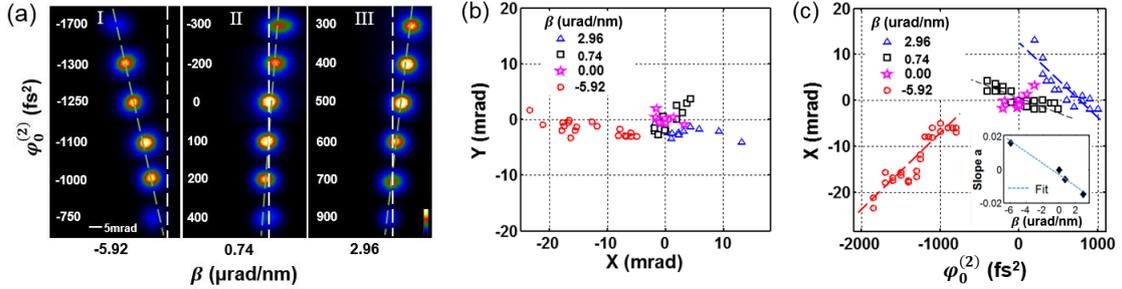

**FIG. 3.** Angular deviation of electrons changed with GDD $\varphi_0^{(2)}$ for different AC $\beta$: (a) Deviations of electron bunches captured on CCD. The white dashed line is the initial optical axis and the green dashed line is the center-to-center connector of electron beams. (b) Horizontal vs. vertical deviation. (c) Horizontal deviation vs. $\varphi_0^{(2)}$. The insert shows the linear fitting of $\beta$ and slope $a$.

The pointing deviations of electron beams were very sensitive to GDD $\varphi_0^{(2)}$ for different AC $\beta$, as shown in Fig. 3. The deviation direction is in the plane of laser polarization, and no obvious deflection was observed in the direction perpendicular to the laser polarization. There is an apparent difference of angular deviation of electron beams for different $\beta$. For a certain $\beta$, the electrons cannot be generated if $\varphi_0^{(2)}$ goes beyond the particular range. The highest angular deviation of electrons is more than 20 mrad when $\beta$ = -5.92 μrad/nm. Figure 3(c) further illustrates the linear relationship between the net deviation of electrons from its center position and $\varphi_0^{(2)}$ under certain $\beta$, which is in accordance with Eq. (1). We use the linear fitting relationship $X[\text{mrad}] = a\,\varphi_0^{(2)}[\text{fs}^2] + b$ to show the linear dependence of electron beam deviations on $\varphi_0^{(2)}$. For $\beta$ = -5.92, 0.74 and 2.96 μrad/nm, the slopes ($a$) are $1.64\times10^{-2}$, $-5.61\times10^{-3}$ and $-1.48\times10^{-2}$ respectively. While $\beta$ = 0.00 μrad/nm, then the PFT $p=0$ and the deviations of the accelerated electrons almost are within 1 mrad and the slope $a$ is negligible. The insert in Fig. 3(c) depicts the linear relationship between $\beta$ and the slope $a$. Actually the physical quantity $a$ represents AC, which will induce an unwanted PFT to the pulse. The laser intensity profile that is tilted with respect to its phase front causes the transverse plasma density gradient. And the transverse electric field pushes the asymmetric wakefield as well as the injected electrons towards different direction. However, the pointing stability of electron bunches has not been affected. Therefore, the pointing-directions of electron-pulses are able to be predictable and guided by controlling AC and GDD.

When $p=0$, the stable electron beams have been obtained and the typical energy distribution recorded on the imaging plate (IP) is shown in Fig. 4(a). The average electron charge is ∼25 pC with energy above 40 MeV and the electron spectrum after spectrometer also stays relatively stable which can be accelerated to ∼180 MeV. The high energy electrons are critical to generate the gamma-ray pulses for ultrafast nuclear dynamics research [24]. Thus, our method is unique and indispensable for steering the high-energy wakefield-accelerated electron beams.

The PFT also plays an important role on the qualities of accelerated electron, including the charge, peak intensity, horizontal and vertical divergence of electron bunches for different $β$, as depicted in Fig. 4(b-e). The electron beams are monitored by a phosphor screen which is coupled with a 16-bit charged-coupled device (CCD) camera. For $β$ =-5.92, 0.74 and 2.96 μrad/nm, the optimum $\varphi_0^{(2)}$ of electron beams are -1100, 50 and 500 $fs^2$ respectively. It indicates that the optimal $\varphi_0^{(2)}$ for each rotation angle of grating is very different. Furthermore, it is easy to find that for the optimum $\varphi_0^{(2)}$, the qualities of electron bunches including the electron energy, are very similar. The small misalignment of grating induces a slight temporal stretching $-2\pi\lambda\beta^2 z/\omega^2$ and the pulse duration will increase from $τ_0$. Meanwhile, the laser spot size in the transverse direction will also increase and result in the laser intensity decreasing [15]. It is a trade-off between lower intensity and longer pulse beam. This could be corrected using a Dazzler system to modify GDD.

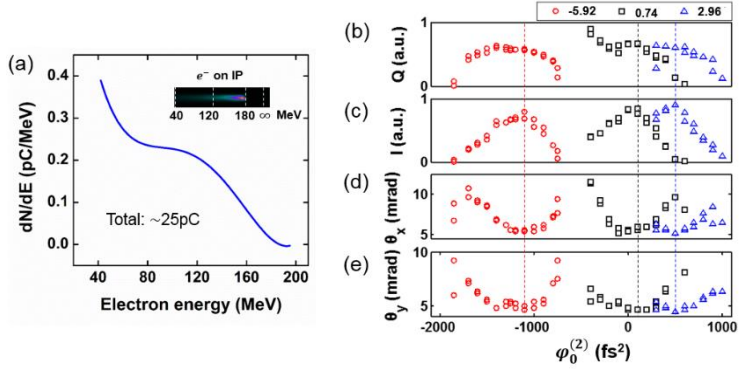

**FIG. 4.** (a) Typical electron spectrum recorded on the imaging plate (IP). The insert is the electron signal on IP after the magnet of 0.9T. Effect of on the qualities of electron beams for different $β$: (b) charge, (c) peak intensity, (d) horizontal and (e) vertical divergence. Here the peak intensity represents charge density. Three dashed lines represent the optimum electron bunch when the pulse duration is corrected with Dazzler system.

This unique method can be utilized to continuously guide the electron-beam pointing-direction to the requisite trajectories, which is very useful and important for practical use. For example, either the all-optical inverse Compton scattering [25,26] or the laser-driven electron-positron collider [8] points out very high requirements on the overlapping between electron beam and colliding beam in space and time. The micrometer-scale deviation will drastically reduce the effectiveness of the physical interaction. This technique offers attractive prospects for driving a bunch of electrons to the colliding point precisely.

More importantly, the magnitude of electron beam's deviation is one order larger than the angular misalignment of gratings. Only if the compressor gratings are strictly parallel to each other, the electron beams will not be deflected from its centers once the PFT is changed. The high degree of sensitivity between electron beam's pointing-direction and PFT provides a precise tool to adjust the parallelism between the compressor gratings of chirped-pulse-amplification system.

### 4. Conclusions

In summary, we present an effective method to precisely steer the high-energy LWFA electron beams with laser GDD. The effect of PFT caused by angular dispersion and temporal change of femtosecond pulses in misaligned stretcher-compressors on the LWFA-electron pointing directions and qualities have been systematically investigated. By using Dazzler system under tilting compressor grating, the electron-pointing-directions can be manipulated. This method is vital to guide the electrons to the requisite trajectories for practical applications. The Dazzler is used to control the specific spectral phase of laser pulses and wouldn't change the initial light path. It is very convenient for practical experiments. It is also the basic principle to judge and adjust the parallelism of the grating compressor in vacuum. This provides a new way to precisely diagnose the spatio-temporal distorted pulses with PFT.

**Funding**

This work was supported by the National Key R&D Program of China (2017YFA0403301), National Natural Science Foundation of China (11334013, 11721404, 11805266), the Key Program of CAS (XDB17030500, XDB16010200) and Science Challenge Project (TZ2018005).

**Disclosures**

The authors declare no conflicts of interest.

**References**

1. T. Tajima and J. M. Dawson, "Laser Electron Accelerator," Phys. Rev. Lett. **43**(4), 267–270 (1979).
2. D. Strickland and G. Mourou, "Compression of amplified chirped optical pulses," Optics communications, **56**(3), 219-221 (1985).
3. E. Esarey, C. B. Schroeder, and W. P. Leemans, "Physics of laser-driven plasma-based electron accelerators," Rev. Mod. Phys. **81** (3), 1229-1285 (2009).
4. X. Wang, R. Zgadzaj, N. Fazel, Z. Li, S. A. Yi, X. Zhang, W. Henderson, Y. Y. Chang, R. Korzekwa, H. E. Tsai, C. H. Pai, H. Quevedo, G. Dyer, E. Gaul, M. Martinez, A. C. Bernstein, T. Borger, M. Spinks, M. Donovan, V. Khudik, G. Shvets, T. Ditmire, and M. C. Downer, "Quasi-monoenergetic laser-plasma acceleration of electrons to 2 GeV," Nat. commun. **4**, 1988 (2013).
5. W. P. Leemans, A. J. Gonsalves, H. S. Mao, K. Nakamura, C. Benedetti, C. B. Schroeder, C. Toth, J. Daniels, D. E. Mittelberger, S. S. Bulanov, J. L. Vay, C. G. Geddes, and E. Esarey, "Multi-GeV electron beams from capillary-discharge-guided subpetawatt laser pulses in the self-trapping regime," Phys. Rev. Lett. **113** (24), 245002 (2014).
6. Y. F. Li, D. Z. Li, K. Huang, M. Z. Tao, M. H. Li, J. R. Zhao, Y. Ma, X. Guo, J. G. Wang, M. Chen, N. Hafz, J. Zhang, and L. M. Chen, "Generation of 20 kA electron beam from a laser wakefield accelerator," Phys. Plasmas **24** (2), 023108 (2017).
7. J. P. Couperus, R. Pausch, A. Kohler, O. Zarini, J. M. Kramer, M. Garten, A. Huebl, R. Gebhardt, U. Helbig, S. Bock, K. Zeil, A. Debus, M. Bussmann, U. Schramm, and A. Irman, "Demonstration of a beam loaded nanocoulomb-class laser wakefield accelerator," Nat. commun. **8** (1), 487 (2017).
8. W. Leemans and E. Esarey, "Laser-driven plasma-wave electron accelerators," Physics Today **62** (3), 44-49 (2009).
9. M. Fuchs, R. Weingartner, A. Popp, Z. Major, S. Becker, J. Osterhoff, I. Cortrie, B. Zeitler, R. Hörlein, G. D. Tsakiris, U. Schramm, T. P. Rowlands-Rees, S. M. Hooker, D. Habs, F. Krausz, S. Karsch, and F. Grüner, "Laser-driven soft-X-ray undulator source," Nat. Physics **5** (11), 826-829 (2009).
10. A. R. Maier, A. Meseck, S. Reiche, C. B. Schroeder, T. Seggebrock, and F. Grüner, "Demonstration Scheme for a Laser-Plasma-Driven Free-Electron Laser," Physical Review X **2** (3) (2012).
11. A. Popp, J. Vieira, J. Osterhoff, Z. Major, R. Horlein, M. Fuchs, R. Weingartner, T. P. Rowlands-Rees, M. Marti, R. A. Fonseca, S. F. Martins, L. O. Silva, S. M. Hooker, F. Krausz, F. Gruner, and S. Karsch, "All-Optical Steering of Laser-Wakefield-Accelerated Electron Beams," Phys. Rev. Lett. **105** (21) (2010).
12. N. Nakanii, T. Hosokai, K. Iwasa, S. Masuda, A. Zhidkov, N. Pathak, H. Nakahara, Y. Mizuta, N. Takeguchi, R. Kodama, "Transient magnetized plasma as an optical element for high power laser pulses," Physical Review Special Topics - Accelerators and Beams **18** (2) (2015).
13. W. Lu, M. Tzoufras, C. Joshi, F. S. Tsung, W. B. Mori, J. Vieira, R. A. Fonseca, and L. O. Silva, "Generating multi-GeV electron bunches using single stage laser wakefield acceleration in a 3D nonlinear regime," Physical Review Special Topics - Accelerators and Beams **10** (6) (2007).
14. S. Akturk, X. Gu, Erik Zeek, and Rick Trebino, "Pulse-front tilt caused by spatial and temporal chirp," Optics express **12** (19), 4399-4410 (2004).
15. G. Pretzler, A. Kasper, and K.J. Witte, "Angular chirp and tilted light pulses in CPA lasers," Applied Physics B **70**, 1-9 (2000).
16. L. J. Wong and I. Kaminer, "Ultrashort Tilted-Pulse-Front Pulses and Nonparaxial Tilted-Phase-Front Beams," ACS Photonics **4** (9), 2257-2264 (2017).
17. K. Poder R. J. Garland, J. Cole, W. Schumaker, D. Doria, L. A. Gizzi, G. Grittani, K. Krushelnick, S. Kuschel, S. P. D. Mangles, Z. Najmudin, D. Symes, A. G. R. Thomas, M. Vargas, M. Zepf, and G. Sarri, "Optimisation of the pointing stability of laser-wakefield accelerated electron beams," arXiv:1407.6979v1 (2014).
18. M. Thévenet, D.E. Mittelberger, K. Nakamura, R. Lehe, C.B. Schroeder, J.-L. Vay, E. Esarey, W. P. Leemans, "Pulse front tilt steering in laser plasma accelerators," Physical Review Accelerators and Beams **22** (7) (2019).
19. S. Afhami, E. Eslami, "Injection and acceleration of electron bunch in a plasma wakefield produced by a chirped laser pulse," Phys. Plasmas **21** (6), 063108 (2014).
20. H. T. Kim, V. B. Pathak, K. Hong Pae, A. Lifschitz, F. Sylla, J. H. Shin, C. Hojbota, S. K. Lee, J. H. Sung, H. W. Lee, E. Guillaume, C. Thaury, K. Nakajima, J. Vieira, L. O. Silva, V. Malka, and C. H. Nam, "Stable multi-GeV electron accelerator driven by waveform-controlled PW laser pulses," Sci. Rep. **7** (1), 10203 (2017).
21. C. Zhu, J. Wang, J. Feng, Y. Li, D. Li, M. Li, Y. He, J. Ma, J. Tan, B. Zhang, W. Yan and L. Chen "Inverse Compton scattering X-ray source from laser electron accelerator in pure nitrogen with 15 TW laser pulses," Plasma Phys. Control. Fusion **61** (024001) (2019).
22. P. Tournois, "Acousto-optic programmable dispersive filter for adaptive compensation of group delay time dispersion in laser systems," Optics Commun. **140** (4-6), 245-249 (1997).
23. A. Weiner, " Ultrafast optics wiley series in pure and applied optics," A John Wiley & Sons, Inc., Publication (2009).
24. J. X. Li, K. Z. Hatsagortsyan, B. J. Galow, and C. H. Keitel, "Attosecond Gamma-Ray Pulses via Nonlinear Compton Scattering in the Radiation-Dominated Regime," Phys. Rev. Lett. **115** (20), 204801 (2015).
25. N. D. Powers, I. Ghebregziabher, G. Golovin, C. Liu, S. Chen, S. Banerjee, J. Zhang, and D. P. Umstadter, "Quasi-monoenergetic and tunable X-rays from a laser-driven Compton light source," Nat. Photonics **8** (1), 29-32 (2014).
26. W. Yan, C. Fruhling, G. Golovin, D. Haden, J. Luo, P. Zhang, B. Zhao, J. Zhang, C. Liu, M. Chen, S. Chen, S. Banerjee, D. Umstadter, "High-order multiphoton Thomson scattering," Nat. Photonics **11**, 514-520 (2017).